\documentstyle[12pt]{ioplppt}
\def\bea{\begin{eqnarray}}
\def\eea{\end{eqnarray}}
\def\tfrac#1#2{{\textstyle {#1 \over #2}}}
\def\binom#1#2{{#1 \choose #2}}
\def\qqquad{\qquad \qquad \qquad } 
\eqnobysec

\begin{document}
\jl{1}

\title{6$j$-symbols for symmetric representations of SO($n$) as the 
double series}[6$j$-symbols of SO($n$)]

\author{Sigitas Ali\v sauskas}

\address{Institute of Theoretical Physics and Astronomy of Vilnius
University, \\
A. Go\v stauto 12, Vilnius 2600, Lithuania}

\begin{abstract}
The corrected triple sum expression of Ali\v sauskas (1987) for the 
recoupling (Racah) coefficients (6$j$-symbols) of the symmetric (most 
degenerate) representations of the orthogonal groups SO($n$) (previously 
derived from the fourfold sum expression of Ali\v sauskas also related to 
result of Horme{\ss} and Junker 1999) is rearranged into three new 
different double sum expressions (related to the hypergeometric Kamp\'{e} 
de F\'{e}riet type series) and a new triple sum expression with 
preferable summation condition. The Regge type symmetry of special 
6$j$-symbols of the orthogonal groups SO($n$) in terms of special 
Kamp\'{e} de F\'{e}riet $F_{1:3}^{1:4}$ series is revealed. The 
recoupling coefficients for antisymmetric representations of symplectic 
group Sp($2n$) are derived using their relation with the recoupling 
coefficients of the formal orthogonal group SO($-2n$).
\end{abstract}

%\pacs{02.20.Q, 02.30.G}

\section{Introduction}

The importance of 6$j$ (Racah) coefficients of SU(2) for the quantum 
angular momentum theory is well known, as well as their applications in 
many branches of mathematical physics, representation theory of Lie and 
quantum groups, in theory of orthogonal polynomials and other special 
functions. The Racah coefficients (6$j$-symbols) and other recoupling 
coefficients of the unitary SU($n$), orthogonal SO($n$) and symplectic 
Sp($2n$) groups of different rank are useful when calculating energy 
levels and transition rates in atomic, molecular and nuclear theory (for 
example, in connection with Jahn--Teller effect and structural analysis 
of atomic shells, see many papers of Judd and co-workers [1--7], 
%\cite{JLO86,J87,JLL90,JLS90,JL87,JL89,JLV00},
for description of multi-fermionic systems and in the microscopic nuclear 
theory [8--14]) %\cite{Kr67,MCh68,H75,V71,V81,V88,VC-A92})
and in conformal field theory \cite{PZ01}. 

Special classes of coupling coefficients and 6$j$-symbols of the SO($n$) 
groups were considered by Ali\v sauskas \cite{Al87,Al02b}, Junker and 
Horme{\ss} \cite{Ju93,HJu99}, with the fourfold \cite{Al87,HJu99} and 
triple \cite{Al87} sum expressions for the recoupling coefficients with 
all most degenerate (symmetric or class-one) irreducible representations. 
(Such 6$j$-symbols have application in the statistical physics, in the 
high-temperature expansion of the SO($n$)-symmetric classical lattice 
models [18--21]). %\cite{Ju93,HJu99,Jo67,D72}). 
Other special expressions for 6$j$-symbols of the SO($n$) were also 
considered in [22--25] %\cite{CK82,C87,PD96,RS98} 
and extended to the Racah coefficients of the 
quantum algebras O$_{q}(n)$ \cite{DPD01}.

The fourfold sum expressions (5.1)--(5.3) of \cite{Al87} and (2)--(10) of 
\cite{HJu99} (cf.\ the integral representation in section 6 of 
\cite{Ju93}) for the 6$j$-symbols of SO($n$) with all six irreducible 
representations (irreps) symmetric are equivalent, taking into account 
the different expressions (11)--(15) of \cite{HJu99} and 
(3.10$a$)--(3.10$b$) of \cite{Al02b} for the integrals involving triplets 
of the Gegenbauer polynomials in terms of the very well-poised $_7F_6(1)$ 
or balanced $_4F_3(1)$ hypergeometric series, related to the $6j$ 
coefficients of SU(2). The Biedenharn--Elliott identity \cite{B53,E53} 
(see [29--31]), %\cite{JB77,VMK88,BL81}), 
used in two stages and related expansions 
allowed us \cite{Al87} to derive the triple sum expression (5.7) for the 
corresponding 6$j$-symbols of SO($n$). Note that the phase factor 
$(-1)^{(g-e)/2}$ (where $g\geq e$) should be omitted in the right-hand 
side of this expression, in contrast with (5.5) of the same paper.

Expressions (5.3) and (5.7) of \cite{Al87} for the 6$j$-symbols of SO($n$)
are given as expansions in terms of three and two multiplied $6j$
coefficients of SU(2) (with some multiple of 1/4 parameters for odd $n$),
respectively. The corresponding sums over the angular momentum type 
parameters resemble the usual expansions \cite{JB77,VMK88} of $9j$ and 
$12j$ coefficients of SU(2) in terms of $6j$ coefficients, which recently 
were rearranged by Rosengren \cite{R99} (for the SU(1,1) group) and 
Ali\v sauskas [33--35] %\cite{Al99,Al00,Al02a} 
using the appropriate (less symmetric)
expressions (29.1$b$) and (29.1$c$) of Jucys and Bandzaitis \cite{JB77} 
(see also (5) and (6) in section 9.2 of \cite{VMK88}) for the Racah 
coefficients (related to the balanced hypergeometric $_4F_3(1)$ 
series) and Dougall's summation formula \cite{Sl66} of the very 
well-poised $_4F_3(-1)$ series. In \cite{Al99,Al02a}, Dougall's 
summation formula \cite{Sl66,GR90} of the very well-poised hypergeometric 
$_5F_4(1)$ series, together with the corresponding expressions for 
the Racah coefficients, was suitable for rearrangement of $12j$ 
coefficients of SU(2). This way the total number of sums in expressions 
was reduced.

In this paper, the triple sum expression (5.7) of \cite{Al87} for the 
6$j$-symbols of SO($n$) with all six irreps symmetric is rearranged in 
a similar manner into the different double sum expressions of the 
hypergeometric (Kamp\'{e} de F\'{e}riet \cite{K-F21,AK-F26}) type, as 
well as into the triple sum expression, with all three separate sums of 
the balanced $_4F_3(1)$ type restricted by a single parameter.

In section 2, the main results of \cite{Al87} concerning the $6j$-symbols 
of SO($n$) are summarized and reconsidered in view of our objectives and
some approaches used in [33--35] %\cite{Al99,Al00,Al02a} 
in the case of $9j$ and 
$12j$ coefficients of SU(2). Three new double sum expressions for the 
renormalized $6j$-symbols of SO($n$) (specified in terms of so-called 
$\alpha $-graphs $I_{n}(a,b,e|d,c,f)$ or related rational 
$c_{a,b,e;d,c,f}^{(\alpha ,n)}$ functions of \cite{HJu99}) are derived in 
section 3, where the Regge \cite{R59} type symmetry is also revealed (for 
$n\geq 5$), as well as the role of the Bargmann--Shelepin \cite{B62,Sh64} 
parameters, extended from the $6j$ coefficients of SU(2) or SO(3). Triple 
sum expression for the renormalized $6j$-symbols of SO($n$) presented in 
section 4 sometimes may be more preferable, similarly as special 
expressions of the stretched or almost stretched $6j$-symbols of SO($n$).

In section 5, the renormalized $6j$-symbols of SO($n$) are expanded in 
terms of (numerator) Pochhammer symbols, as well as in terms of special 
class of Kamp\'{e} de F\'{e}riet \cite{K-F21,AK-F26}) functions 
$F_{1:3;3}^{1:4;4}[{{...}\atop {...}};1,1]$, which specific features and 
diversity are considered.

The recoupling coefficients for antisymmetric representations 
$\langle 1^{\nu }\rangle $ of symplectic group Sp($2n$) are presented in 
appendix as formal analytical continuation of the recoupling coefficients 
for symmetric representations of the orthogonal group with negative rank 
SO($-2n$), in accordance with \cite{JLL90,JLV00,CK82} (cf.\ also 
\cite{M80,D89}).

\section{Preliminaries}

In accordance with (5.3) of \cite{Al87}, we may express the $6j$-symbol 
of SO($n$) ($n\geq 4$) with all representations symmetric as follows: 
\bea
\fl \left\{ \begin{array}{ccc}
a & b & e \\ 
d & c & f
\end{array} \right\} _{SO(n)} =\left[ \frac{(2c+n-2)(2d+n-2)(2e+n-2)}{%
8\,d_{c}^{(n)}d_{d}^{(n)}d_{e}^{(n)}}\right] ^{1/2}  \nonumber \\
\times \left( \begin{array}{ccc}
c & d & e \\ 
0 & 0 & 0
\end{array} 
\right) _{n}^{-1}\sum_{l^{\prime }}(-1)^{(c+d-e)/2+f+n+l^{\prime }}
(2l^{\prime }+n-3)  \nonumber \\
\times \left\{ \begin{array}{ccc}
\frac 12 b & \frac 12 f+\frac 14 n-1 & \frac 12 d+\frac 14 n-1 \\ 
\frac 12 f+\frac 14 n-1 & \frac 12(b+n)-2 & l^{\prime }+\frac 12 n-2
\end{array} \right\}  \nonumber \\
\times \left\{ \begin{array}{ccc}
\frac 12 a & \frac 12 f+\frac 14 n-1 & \frac 12 c+\frac 14 n-1 \\ 
\frac 12 f+\frac 14 n-1 & \frac 12 (a+n)-2 & l^{\prime }+\frac 12 n-2
\end{array} \right\}  \nonumber \\
\times \left\{ \begin{array}{ccc}
\frac 12 a & \frac 12 b+\frac 14 n-1 & \frac 12 e+\frac 14 n-1 \\ 
\frac 12 b+\frac 14 n-1 & \frac 12 (a+n)-2 & l^{\prime }+\frac 12 n-2
\end{array} \right\} \left[ \frac{l^{\prime }!(n-3)!}{(l^{\prime }
+n-4)!}\right] ^{1/2},  \label{df6j}
\eea
where in the right-hand side the usual $6j$ coefficients of SU(2) [29--31]
%\cite{JB77,VMK88,BL81} 
appear for $n$ even. Otherwise, for $n$ odd some 
integer linear combination of parameters of the type $a-l^{\prime }$, 
$l^{\prime }$ or $(b+d-f)/2$ are also restricting the summation intervals 
in extensions of the asymmetric (Jucys--Bandzaitis) expressions for $6j$ 
coefficients (as presented by (2.1$a,b$) and (2.2$a,b$) in \cite{Al99} or 
\cite{Al00} for $q=1$), with some ratios of factorials $x!/y!$ 
turning into ratios of the gamma functions $\Gamma (x+1)/\Gamma (y+1)$ 
with half-integer arguments.

The dimension 
\begin{equation}
d_{l}^{(n)}=\frac{(2l+n-2)(l+n-3)!}{l!(n-2)!}  \label{dimo}
\end{equation}
of the SO($n$) symmetric irreducible representation $l$ and special 
$3j$-symbols 
\numparts \bea
\fl \left( \begin{array}{ccc}
l_1 & l_2 & l_{3} \\ 
0 & 0 & 0
\end{array}
\right) _{n}=(-1)^{\psi _n}\frac{1}{\Gamma (n/2)}\left[ \frac{(J+n-3)!}{%
(n-3)!\Gamma (J+n/2)}\right.  \nonumber \\
\times \left. \prod_{i=1}^{3}\frac{\left( l_i+n/2-1\right) \Gamma 
\left( J-l_i+n/2-1\right) }{d_{l_i}^{(n)}(J-l_i)!}\right] ^{1/2}  
\label{isf0} \\
\lo= (-1)^{\psi _n}\frac{\widetilde{\nabla }_{n[0,1,2,3]}^{-1}(l_1,l_2,
l_3)}{\Gamma (n/2)}\left[ \frac{1}{(n-3)!}\prod_{i=1}^{3}\frac{l_i+n/2
-1}{d_{l_i}^{(n)}}\right] ^{1/2}  \label{isf0n}
\eea \endnumparts
(see [16--18] %\cite{Al87,Al02b,Ju93} 
and related special Clebsch--Gordan 
coefficients \cite{KK73,NAl74a}), used in (\ref{df6j}) and further, are 
rational numbers (in part under the square root sign). In (\ref{isf0}), 
$J=\frac 12(l_1+l_2+l_{3})$ and $J-l_i$ ($i=1,2,3$) are non-negative 
integers. The triangular coefficient $\widetilde{\nabla }_{n[0,1,2,3]}
(l_1,l_2,l_{3})$ in (\ref{isf0n}) may be expressed as follows 
\bea
\fl \widetilde{\nabla }_{n[0,1,2,3]}(a,b,e)=\left( \frac{\left[ 
\frac 12(b+e-a)\right] !\left[ \frac 12 (a-b+e)\right] !}{\Gamma \left( 
\frac 12(b+e-a+n)-1\right) \Gamma \left( \frac 12 (a-b+e+n)-1\right) }
\right.  \nonumber \\
\times \left. \frac{\left[ \frac 12 (a+b-e)\right] !\Gamma \left( 
\frac12(a+b+e+n)\right) }{\Gamma \left( \frac 12(a+b-e+n)-1\right) \left[ 
\frac 12(a+b+e)+n-3\right] !}\right) ^{1/2}.  \label{nabls}
\eea
It is reasonable to take $\psi _n=0$ for $n\geq 4$ (cf.\ \cite
{Al87,Al02b,NAl74a}).

Six (from 24) elementary symmetry properties of the $6j$-symbols of 
SO($n$) 
\begin{equation}
\fl \left\{ \begin{array}{ccc}
j_1 & j_2 & j_3 \\ 
l_1 & l_2 & l_3
\end{array} \right\} _{SO(n)}=\left\{ \begin{array}{ccc}
j_a & j_b & j_c \\ 
l_a & l_b & l_c
\end{array} \right\} _{SO(n)}=\left\{ \begin{array}{ccc}
j_a & l_b & l_c \\ 
l_a & j_b & j_c
\end{array} \right\} _{SO(n)}  \label{sym6j}
\end{equation}
are visible from expression (\ref{df6j}) (see also (16) of \cite{HJu99}).

We may replace the last two factors (in the last line) on the right-hand 
side of (\ref{df6j}) by 
\bea
\fl \widetilde{\nabla }_{n[0,3,5,6]}(a,b;e,0)\sum_{g\geq e}\frac{(g+n-3)
\Gamma \left( \frac 12(g-e+n)-2\right) \left[ \frac 12 (g+e)+n-4\right] !%
}{\left[ \frac 12 (g-e)\right] !\Gamma \left( \frac 12(g+e+n)\right) 
\Gamma \left( \frac 12 n-2\right) }  \nonumber
\\
\times (-1)^{(g-e)/2}\left( \frac{(n-3)!\left[ \frac 12(a-b+g)\right] !
\left[ \frac 12(b-a+g)\right] !}{\left[ \frac 12(a-b+g)+n-4\right] !
\left[ \frac 12(b-a+g)+n-4\right] !}\right) ^{1/2}  \nonumber \\
\times \left\{ \begin{array}{ccc}
\frac 12 (a+n)-2 & \frac 12 a & l^{\prime }+\frac 12 n-2 \\ 
\frac 12 (b+n)-2 & \frac 12 b & \frac 12 (g+n)-2
\end{array} \right\} ,  \label{rpl}
\eea
valid also in accordance with $q=1$ version of expression (2.1$b$) of 
\cite{Al99,Al00} and Dougall's summation formula (2.3.4.5) of \cite{Sl66} 
for special very well-poised $_5F_4(1)$ series as presented by an 
extension of (A1$a$) of \cite{Al02a} ,or (A4$a$) of \cite{Al99} (after 
replacing $\Gamma (-x)/\Gamma (-y)$ if necessary by 
$(-1)^{x-y}\Gamma (y+1)/\Gamma (x+1)$ for $x-y$ integer) with parameters 
\begin{eqnarray*}
j \rightarrow \tfrac 12(g+n)-2,\qquad p_1\rightarrow -\tfrac 12(e+n)
+1,\qquad p_2\rightarrow \tfrac 12 e, \\
p_3 \rightarrow -\tfrac 12(a+b+n),\qquad p_4\rightarrow \tfrac 12
(a+b+n)-2-l^{\prime }+z
\end{eqnarray*}
and integer $p_1+p_4+1$, restricting the summation interval of 
$_5F_4(1)$ series. Another triangular coefficient 
\bea
\fl \widetilde{\nabla }_{n[0,3,5,6]}(a,b;e,0)=\left( \frac{\Gamma \left( 
\frac 12(b+e-a+n)-1\right) \Gamma \left( \frac 12 (a-b+e+n)-1\right) }{%
\left[ \frac 12 (b+e-a)\right] !\left[ \frac 12 (a-b+e)\right] !}\right.
\nonumber \\
\times \left. \frac{\left[ \frac 12 (a+b-e)\right] !\Gamma \left( \frac 12
(a+b+e+n)\right) }{\Gamma \left( \frac 12(a+b-e+n)-1\right) \left[ 
\frac 12(a+b+e)+n-3\right] !}\right) ^{1/2}.  \label{nabln}
\eea
in (\ref{rpl}) is related to $\nabla _{n[0,3,5,6]}(a,b;e,0)$ as defined 
by (2.3) of \cite{Al87}, but coincides with it only for even $n$.

Hence, the Biedenharn--Elliott identity, applied to triplet of 
6$j$-coefficients of SU(2) in (\ref{df6j}) with substituted (\ref{rpl}), 
allowed us to present special $6j$-symbol of SO($n$) (cf.\ (5.7) of 
\cite{Al87}) for $n>4$ as follows: 
\bea
\fl \left\{ \begin{array}{ccc}
a & b & e \\ 
d & c & f
\end{array}
\right\} _{SO(n)}=\left[ \frac{(2c+n-2)(2d+n-2)(2e+n-2)}{%
8\,d_c^{(n)}d_d^{(n)}d_e^{(n)}}\right] ^{1/2}\left( 
\begin{array}{ccc}
c & d & e \\ 
0 & 0 & 0
\end{array} \right) _n^{-1}  \nonumber \\
\times \widetilde{\nabla }_{n[0,3,5,6]}(a,b;e,0)\sum_{g=e}^{a+b}
\frac{(g+n-3)\,\Gamma \left( \frac 12 (g-e+n)-2\right) }{\left[ \frac 12 
(g-e)\right] !\Gamma \left( \frac 12 (g+e+n)\right) \Gamma \left( 
\frac 12 n-2\right) }  \nonumber \\
\times \left( \frac{\left[ \frac 12(a-b+g)\right] !\left[ \frac 12(b-a+g)
\right] !(n-3)!}{\left[ \frac 12(a-b+g)+n-4\right] !\left[ \frac 12
(b-a+g)+n-4\right] !}\right) ^{1/2}  \nonumber \\
\times \left[ \tfrac 12(g+e)+n-4\right] !\left\{ \begin{array}{ccc}
\frac 12 c+\frac 14 n-1 & \frac 12 a & \frac 12 f+\frac 14 n-1 \\ 
\frac 12(b+n)-2 & \frac 12 d+\frac 14 n-1 & \frac 12(g+n)-2
\end{array} \right\}  \nonumber \\
\times \left\{ \begin{array}{ccc}
\frac 12 b & \frac 12(a+n)-2 & \frac 12 (g+n)-2 \\ 
\frac 12 c+\frac 14 n-1 & \frac 12 d+\frac 14 n-1 & \frac 12 f+\frac 14 
n-1
\end{array} \right\} ,  \label{rtr6j}
\eea
with product of two 6$j$-coefficients of SU(2) (some parameters of which
accept values multiple of 1/4 for odd $n$) in the right-hand side. It was 
suggested in \cite{Al87} to use for them the most symmetric (Racah) 
expression (see [29--31]) %\cite{JB77,VMK88,BL81}) which is
useless, however, for rearrangement of (\ref{rtr6j}).

\section{Double sum expressions for $6j$-symbols of SO($n$)}

Nevertheless, we may rearrange (\ref{rtr6j}) expressing the second 
6$j$-coefficient of SU(2) by means of (2.1a) of \cite{Al99,Al00} and the 
first one by means of (2.1b) of \cite{Al99,Al00}. In this case the factors, 
depending on the summation parameter $j=(g+n)/2-2$ and distributed in the 
numerators and denominators of different 6$j$-coefficients under the 
square root, cancel, together with the asymmetric triangle coefficients 
\numparts \bea
\fl \nabla (xyj)=\left[ \frac{(x+y-j)!(x-y+j)!(x+y+j+1)!}{(y+j-x)!}
\right] ^{1/2}  \label{nabla} \\
\lo= \left[ \frac{\Gamma (x+y-j+1)\Gamma (x-y+j+1)\Gamma (x+y+j+2)}{%
\Gamma (y+j-x+1)}\right] ^{1/2}.  \label{nablg}
\eea \endnumparts
Then we again may use the Dougall's summation formula for very 
well-poised $_5F_4(1)$ series as presented by (A1$b$) of \cite{Al02a} 
(see (A4$b$) of \cite{Al99}) with parameters 
\begin{eqnarray*}
j\rightarrow \tfrac 12 (g+n)-2,\qquad p_1\rightarrow -\tfrac 12 
(e+n)+1,\qquad p_2\rightarrow \tfrac 12 e, \\
p_3 \rightarrow \tfrac 12 (f-b-c)-1+z_2,\qquad p_4\rightarrow \tfrac 12
(b+c-f+n)-2+z_1.
\end{eqnarray*}

Two more different rearrangements of (\ref{rtr6j}) are also possible 
in the following ways: The second version may be obtained when we express 
the last 6$j$-coefficients of SU(2) in the right-hand side
\begin{equation}
\left\{ \begin{array}{ccc}
\frac 12(a+n)-2 & \frac 12 f+\frac 14 n-1 & \frac 12 c+\frac 14 n-1 \\ 
\frac 12 d+\frac 14 n-1 & \frac 12(g+n)-2 & \frac 12 b
\end{array} \right\}  \label{ch6ja}
\end{equation}
(with transposed parameters) by means of (2.2$a$) of \cite{Al99,Al00} and 
use extended version of (A1$a$) of \cite{Al02a} (or (A4$a$) of 
\cite{Al99}) with parameters 
\begin{eqnarray*}
j\rightarrow \tfrac 12 (g+n)-2,\qquad p_1\rightarrow -\tfrac 12 %
(e+n)+1,\qquad p_2\rightarrow \tfrac 12 e, \\
p_3\rightarrow z_2-\tfrac 12 (b+c+f+n),\qquad p_4\rightarrow \tfrac 12
(b+c-f+n)-2+z_1.
\end{eqnarray*}
The third version may be obtained when we express the 6$j$-coefficients 
of SU(2) in the right-hand side of (\ref{rtr6j}) 
\bea
\fl \left\{ \begin{array}{ccc}
\frac 12(g+n)-2 & \frac 12 a & \frac 12(b+n)-2 \\ 
\frac 12 f+\frac 14 n-1 & \frac 12 d+\frac 14 n-1 & \frac 12 c+\frac 14 
n-1
\end{array} \right\}  \nonumber \\
\times \left\{ \begin{array}{ccc}
\frac 12 c+\frac 14 n-1 & \frac 12 f+\frac 14 n-1 & \frac 12(a+n)-2 \\ 
\frac 12 b & \frac 12(g+n)-2 & \frac 12 d+\frac 14 n-1
\end{array} \right\}  \label{ch6jb}
\eea
by means of (2.2$a$) and (2.1$a$) of \cite{Al00}, respectively, and use 
an extended version of (A1$b$) of \cite{Al02a} with parameters 
\begin{eqnarray*}
j\rightarrow \tfrac 12(g+n)-2,\qquad p_1\rightarrow -\tfrac 12(e+n)+1,
\qquad p_2\rightarrow \tfrac 12 e, \\
p_3\rightarrow \tfrac 12(c-d)-1-z_2,\qquad p_4\rightarrow \tfrac 12 %
(a+b+n)-2-z_1.
\end{eqnarray*}
In all three cases the summation intervals over $j$ (or $g$) are 
restricted by non-negative integers $p_1+p_4+1$. In contrast with the 
case of $9j$ and $12j$ coefficients of SU(2) (see [33--35]), 
%\cite{Al99,Al00,Al02a}), 
the formal summation intervals over $g$ cannot exceed $\frac 12 
\min (a+b-e,c+d-e)$ (determined by triangular conditions) in main and 
replaced by (\ref{ch6ja}) versions of (\ref{rtr6j}). Nevertheless, the 
possible superfluous terms arising for $g=c+d+2,c+d+4,...$ in the third 
version of (\ref{rtr6j}) (with $6j$ coefficients replaced by 
(\ref{ch6jb})) are unimportant, since in this case the sum over $z_2$ 
turns into 0, in accordance with Karlssons summation formula \cite{GR90} 
(cf.\ section 2 of \cite{Al99,Al00}).

Now it is convenient to write the $6j$-symbol of SO($n$) in terms of 
so-called $\alpha $-graph $I_n(a,b,e|d,c,f)$ or related quantity 
$c_{a,b,e;d,c,f}^{(\alpha ,n)}$ (see \cite{HJu99}) 
\numparts \bea
\fl \left\{ \begin{array}{ccc}
a & b & e \\ 
d & c & f
\end{array}
\right\} _{SO(n)}=c_{a,b,e;d,c,f}^{(\alpha ,n)}\,\left[
d_a^{(n)}d_b^{(n)}d_c^{(n)}d_d^{(n)}d_e^{(n)}d_f^{(n)}
\left( \begin{array}{ccc}
a & b & e \\ 
0 & 0 & 0
\end{array} \right) _n\right.  \nonumber \\
\times \left. \left( \begin{array}{ccc}
a & c & f \\ 
0 & 0 & 0
\end{array} \right) _n\left( \begin{array}{ccc}
b & d & f \\ 
0 & 0 & 0
\end{array} \right) _n\left( \begin{array}{ccc}
c & d & e \\ 
0 & 0 & 0
\end{array} \right) _n\right] ^{-1}  \label{rd6ja} \\
\lo= \widetilde{\nabla }_{n[0,1,2,3]}(a,b,e)\widetilde{\nabla }%
_{n[0,1,2,3]}(a,c,f)\widetilde{\nabla }_{n[0,1,2,3]}(b,d,f)  \nonumber \\
\times \widetilde{\nabla }_{n[0,1,2,3]}(c,d,e)\frac{[(n-3)!]^{2}\Gamma ^4
\left( n/2\right) }{(a,b,c,d,e,f)_{[n]}}c_{a,b,e;d,c,f}^{(\alpha ,n)},
\label{rd6jb}
\eea \endnumparts
where 
\begin{eqnarray*}
\fl (a,b,c,d,e,f)_{[n]}=\tfrac{1}{64}(2a+n-2)(2b+n-2)(2c+n-2) \\
\times (2d+n-2)(2e+n-2)(2f+n-2),
\end{eqnarray*}
and the quantities 
\bea
\fl c_{a,b,e;d,c,f}^{(\alpha ,n)}=d_a^{(n)}d_b^{(n)}d_c^{(n)}d_d^{(n)}
d_e^{(n)}d_f^{(n)}I_n(a,b,e|d,c,f)  \nonumber \\
\lo= d_a^{(n)}d_b^{(n)}d_c^{(n)}d_d^{(n)}d_e^{(n)}d_f^{(n)}\int_{SO(n)}
\d g_1\int_{SO(n)}\d g_2\int_{SO(n)}\d g_3D_{00}^{a}(g_1)  \nonumber \\
\times D_{00}^{b}(g_2)D_{00}^{e}(g_3)D_{00}^{d}(g_2^{-1}g_3)D_{00}^{c}
(g_3^{-1}g_1)D_{00}^{f}(g_1^{-1}g_2)  \label{cgr}
\eea
are rational numbers and the triangular coefficients $\widetilde{\nabla }%
_{n[0,1,2,3]}(l_1,l_2,l_3)$ are defined by (\ref{nabls}). Here 
$D_{00}^{l}(g)$ are the zonal spherical functions \cite{Vi65} of irrep 
$l$ of SO($n$). In our phase system with $\psi _n=0$ the phase factor 
$(-1)^{d+e+f}$ of (2) of \cite{HJu99} vanishes.

From (\ref{rtr6j}) after summation over $g$ we obtain three following 
different expressions for coefficients (\ref{cgr}):
\numparts \bea
\fl c_{a,b,e;d,c,f}^{(\alpha ,n)}=(a,b,c,d,e,f)_{[n]}\frac{\left[ 
\frac 12(a+c+f)+n-3\right] !\Gamma \left( \frac 12 (a+c-f+n)-1\right) }{%
(n-3)!\Gamma \left( \frac 12 (a+c+f+n)\right) \left[ \frac 12(a+c-f)
\right] !}  \nonumber \\
\times \frac{\Gamma \left( \frac 12 (b+e-a+n)-1\right) \Gamma \left( 
\frac 12(a-b+e+n)-1\right) }{\Gamma ^{3}\left( \frac 12 n\right) \left[ 
\frac 12(b+e-a)\right] !\left[ \frac 12 (a-b+e)\right] !}
(-1)^{(b+c-e-f)/2}  \nonumber \\
\times \sum_{z_1,z_2}\frac{(-1)^{z_1+z_2}\Gamma \left( 
\frac 12(b+d-f+n)-1+z_1\right) \left[ \frac 12(a+c-f)+z_1\right] !}{z_1!
\left[ \frac 12 (a-c+f)-z_1\right] !\left[ \frac 12(d+f-b)-z_1\right] !} 
\nonumber \\
\times \frac{\Gamma \left( f+\frac 12 n-1-z_1\right) \Gamma 
\left( \frac 12 (b+c+e-f+n)-1-z_2\right) }{\left[ \frac 12 (b+c-e-f)
+z_1\right] !\Gamma \left( \frac 12(b+c+e-f+n)+z_1\right) z_2!}  
\nonumber \\
\times \frac{\Gamma \left( \frac 12 (d+f-b+n)-1+z_2\right) \Gamma
\left( \frac 12 (a-c+f+n)-1+z_2\right) }{\left[ \frac 12 (b+d-f)
-z_2\right] !\Gamma \left( \frac 12 (a+c-f+n)-1-z_2\right) }  \nonumber \\
\times \frac{(z_1+z_2)!}{\left[ \frac 12 (e+f-b-c)+z_2\right] !
\Gamma \left( f+\frac 12 n+z_2\right) \Gamma \left( \frac 12 n-1+z_1
+z_2\right) }  \label{rinva} \\
\lo= (a,b,c,d,e,f)_{[n]}\frac{\left[ \frac 12 (a+c+f)+n-3\right] !\Gamma
\left( \frac 12 (a+c-f+n)-1\right) }{(n-3)!\Gamma \left( \frac 12
(a+c+f+n)\right) \left[ \frac 12 (a+c-f)\right] !}  \nonumber \\
\times \frac{\Gamma \left( \frac 12 (b+e-a+n)-1\right) \Gamma \left( 
\frac 12(a-b+e+n)-1\right) }{\Gamma ^{3}\left( \frac 12 n\right) \left[ 
\frac 12(b+e-a)\right] !\left[ \frac 12(a-b+e)\right] !}
(-1)^{(a-b-c+d)/2}  \nonumber \\
\times \sum_{z_1,z_2}\frac{(-1)^{z_1+z_2}\Gamma \left( \frac 12(b+d-f+n)
-1+z_1\right) \left[ \frac 12(a+c-f)+z_1\right] !}{z_1!z_2!\left[ \frac 12
(a-c+f)-z_1\right] !\left[ \frac 12 (b+c-e-f)+z_1\right] !}  \nonumber \\
\times \frac{\Gamma \left( f+\frac 12 n-1-z_1\right) (f-z_1-z_2)!
}{\left[ \frac 12(d+f-b)-z_1\right] !\Gamma \left( \frac 12(b+c+e-f+n)
+z_1\right) }  \nonumber \\
\times \frac{\Gamma \left( \frac 12 (b+c-e+f+n)-1-z_2\right) }{\Gamma 
\left( f+\frac 12 n-1-z_1-z_2\right) \left[ \frac 12(b-d+f)
-z_2\right] !\left[ \frac 12(c+f-a)-z_2\right] !}  \nonumber \\
\times \frac{\Gamma \left( f+\frac 12 n-1-z_2\right) \left[ \frac 12
(b+c+e+f)+n-3-z_2\right] !}{\Gamma \left( \frac 12 (b+d+f+n)-z_2\right) 
\left[ \frac 12(a+c+f)+n-3-z_2\right] !}  \label{rinvb} \\
\lo= (a,b,c,d,e,f)_{[n]}\frac{\Gamma \left( \frac 12(c+f-a+n)-1\right)
\Gamma \left( \frac 12(a-c+f+n)-1\right) }{(n-3)!\left[ \frac 12(c+f-a)
\right] !\left[ \frac 12 (a-c+f)\right] !}  \nonumber \\
\times \frac{\Gamma \left( \frac 12 (b+e-a+n)-1\right) \Gamma \left( 
\frac 12(a-b+e+n)-1\right) }{\Gamma ^{3}\left( \frac 12 n\right) \left[ 
\frac 12(b+e-a)\right] !\left[ \frac 12(a-b+e)\right] !}%
(-1)^{(a+d-e-f)/2}  \nonumber \\
\times \sum_{z_1,z_2}\frac{(-1)^{z_1+z_2}\Gamma \left( \frac 12(a+b+c-d+n)
-1-z_1\right) (a-z_1)!}{z_1!z_2!\left[ \frac 12(a+b-e)-z_1\right] !
\Gamma \left( \frac 12(a+b+e+n)-z_1\right) }  \nonumber \\
\times \frac{\left[ \frac 12(a+b+c+d)+n-3-z_1\right] !\Gamma \left( 
\frac 12(d+f-b+n)-1+z_2\right) }{\left[ \frac 12(a+c-f)-z_1\right] !
\Gamma \left( \frac 12(a+c+f+n)-z_1\right) \left[ \frac 12(b-d+f)
-z_2\right] !}  \nonumber \\
\times \frac{\Gamma \left( \frac 12(d+e-c+n)-1+z_2\right) }{\left[ 
\frac 12(c-d+e)-z_2\right] !\Gamma \left( \frac 12(a-b-c+d+n)-1
+z_2\right) }  \nonumber \\
\times \frac{\Gamma \left( \frac 12 (a+b+c-d+n)-1-z_2\right) \left[ 
\frac 12(a+b+c-d)-z_1-z_2\right] !}{\Gamma \left( d+\frac 12 n+z_2\right) 
\Gamma \left( \frac 12(a+b+c-d+n)-1-z_1-z_2\right) },  \label{rinvc}
\eea \endnumparts
without the visible symmetry properties of 6$j$-symbols of the orthogonal 
SO($n$) group. These expressions are valid for $n\geq 4$ (and probably 
for $c_{a,b,e;d,c,f}^{(\alpha ,3)}$). When $n=4$, the numerator and 
denominator factorials (gamma functions) depending on $z_1+z_2$ cancel 
and 6$j$-symbols of SO(4) split into product (in this case equal to the 
square) of two $6j$ coefficients of SU(2).

All separate sums over $z_1$ or $z_2$ in (\ref{rinva})--(\ref{rinvc})
correspond to the terminating balanced (Saalsch\"{u}tzian) $_5F_4(1)$
series (cf.\ \cite{Sl66,GR90}), with summation intervals restricted by 
\numparts \begin{equation}
\fl \tfrac 12\min (a-c+f,d+f-b)\;{\rm and}\;\tfrac 12(d+e-c)\;{\rm for}\;
\tfrac 12(b+c-e-f)\geq 0  \label{sinta}
\end{equation}
or by 
\begin{equation}
\fl \tfrac 12\min (a+b-e,c+d-e)\;{\rm and}\;\tfrac 12(b+d-f)\;{\rm for}\;
\tfrac 12(b+c-e-f)\leq 0  \label{sintb}
\end{equation}
in (\ref{rinva}), by 
\begin{equation}
\fl \tfrac 12\min (a-c+f,d+f-b,a+b-e,c+d-e)\;{\rm and}\;\tfrac 12\min
(b-d+f,c+f-a)  \label{sintc}
\end{equation}
in (\ref{rinvb}) and by 
\begin{equation}
\fl \tfrac 12\min (a+b-e,a+c-f)\;{\rm and}\;\tfrac 12\min (b-d+f,c-d+e)
\label{sinte}
\end{equation} \endnumparts
in (\ref{rinvc}).

Using (\ref{rd6jb}), together with expression (\ref{rinva}) or 
(\ref{rinvb}) for coefficients $c_{a,b,e;d,c,f}^{(\alpha ,n)}$ (after 
cancelling the dimensions of irreps and the $(a,b,c,d,e,f)_{[n]}$ type 
factors), the Regge type symmetry (cf.\ \cite{R59}) 
\begin{equation}
\left\{ \begin{array}{ccc}
a & b & e \\ 
d & c & f
\end{array} \right\} _{SO(n)}=\left\{ \begin{array}{ccc}
s_3-a & s_3-b & e \\ 
s_3-d & s_3-c & f
\end{array} \right\} _{SO(n)}  \label{sym6Ra}
\end{equation}
of special 6$j$-symbols of SO($n$) (where $s_3=\tfrac 12(a+b+c+d)$) may
be checked. Otherwise, using (\ref{rd6jb}), together with expression 
(\ref{rinvc}), the usual and Regge type symmetries 
\numparts \bea
\fl \left\{ \begin{array}{ccc}
a & b & e \\ 
d & c & f
\end{array}
\right\} _{SO(n)}=\left\{ \begin{array}{ccc}
a & s_1-e & s_1-b \\ 
d & s_1-f & s_1-c
\end{array} \right\} _{SO(n)}  \label{sym6Rb} \\
\lo= \left\{ \begin{array}{ccc}
a & s_1-f & s_1-c \\ 
d & s_1-e & s_1-b
\end{array} \right\} _{SO(n)}=\left\{ \begin{array}{ccc}
a & c & f \\ 
d & b & e
\end{array} \right\} _{SO(n)},  \label{sym6Rc}
\eea \endnumparts
where $s_1=\tfrac 12 (b+c+e+f),$ are visible. The symmetries 
(\ref{sym6Ra}) and (\ref{sym6Rb}) correspond to some column 
transpositions of Shelepin's \cite{Sh64} $4\times 3$ $R$-array 
\numparts \bea
\fl \left\{ \begin{array}{ccc}
a & b & e \\ 
d & c & f
\end{array} \right\}=\left| 
\begin{array}{cccc}
a+b-e & a+c-f & b+d-f & c+d-e \\ 
a-c+f & a-b+e & d+e-c & d+f-b \\ 
b-d+f & c-d+e & b+e-a & c+f-a
\end{array} \right|  \label{raSh} \\
\times \left| 
\begin{array}{cccc}
2r_{11} & 2r_{12} & 2r_{13} & 2r_{14} \\ 
2r_{21} & 2r_{22} & 2r_{23} & 2r_{24} \\ 
2r_{31} & 2r_{32} & 2r_{33} & 2r_{34}
\end{array} \right|  \label{rasht}
\eea \endnumparts
of $6j$ coefficients (cf.\ (29.32) of \cite{JB77} or (12) in section 9.1 
of \cite{VMK88}). Array (\ref{raSh}) (cf.\ also \cite{B62}) is also 
convenient for description of 144 symmetries of 6$j$-symbols of SO($n$) 
under arbitrary transpositions of its columns or rows. Since all the 
entrees of (\ref{raSh}) are even integers for $n\geq 4$, integer 
parameters $r_{ik}=\beta _i-\alpha _{k}$ ($i=1,2,3;\;j=1,2,3,4$) may be 
more convenient, as well as the most symmetric parameterization 
\begin{equation}
\fl \begin{array}{c} 
\alpha _1=\tfrac 12(c+d+e),\;\alpha _2=\tfrac 12(b+d+f),\;\alpha _3=
\tfrac 12(a+c+f),\;\alpha _4=\tfrac 12 (a+b+e), \\ 
\beta _1=\tfrac 12 (a+b+c+d),\;\beta _2=\tfrac 12 (a+d+e+f),\;\beta _3=
\tfrac 12 (b+c+e+f),
\end{array}  \label{sym6p}
\end{equation}
with $\alpha _1+\alpha _2+\alpha _3+\alpha _4=\beta _1+\beta _2+\beta _3$ 
(cf.\ \cite{JB77}).

Expression (\ref{rinva}) includes the minimum of terms, when minimal 
values are accepted by the parameters in the same (the first or the 
second) row of array (\ref{raSh}) ($r_{i3}$ and $r_{i1}$ or $r_{i4}$, for 
$i=1$ or 2). Otherwise, (\ref{rinvb}) and (\ref{rinvc}) include the 
minimum of terms, when minimal values are accepted by the parameters in 
the same column (respectively, $r_{i1}$ and $r_{31}$ or $r_{i4}$ and 
$r_{34}$, $i=1$ or 2 for (\ref{rinvb}), or $r_{11}$ and $r_{31}$ or 
$r_{14}$ and $r_{34}$ in the last case). As a rule (with single exception 
in each case), the definite triplets of numerator and denominator 
factorials or gamma functions, depending on summation parameters $z_1$, 
$z_2$ and $z_1+z_2$, form in (\ref{rinva}), (\ref{rinvb}) and 
(\ref{rinvc}) the binomial coefficients (e.g., $z_1!$, $z_2!$ and 
$(z_1+z_2)!$), their analytical continuation or beta functions which 
respond to relations (\ref{bala}) or (\ref{balb}) between the parameters 
of special Kamp\'{e} de F\'{e}riet \cite{K-F21,AK-F26} functions 
$F_{1:3;3}^{1:4;4}[{{...}\atop {...}};1,1]$, considered in section 5.

\section{Expressions for $6j$-symbols of SO($n$) with summation 
restricted by single parameter}

The double sum expressions of $6j$-symbols of SO($n$) may be inconvenient,
when $r_{11}\ll r_{1k}$ ($k=2,3,4$) and $r_{11}\ll r_{i1}$ ($i=2,3$). In 
the stretched case of $6j$-symbols of SO($n$) (with $r_{11}=0$) we obtain 
\bea
\fl c_{a,b,a+b;d,c,f}^{(\alpha ,n)}=\frac{(a,b,c,d,e,f)_{[n]}\Gamma 
\left( a+\frac 12 n-1\right) \Gamma \left( b+\frac 12 n-1\right) }{(n-3)!
\Gamma ^{3}\left( \frac 12 n\right) \nabla ^{2}\left( \frac 12 a,\frac 12 
c+\frac 14 n-1,\frac 12 f+\frac 14 n-1\right) }  \nonumber \\
\times \frac{\nabla ^{2}\left( \frac 12 (e+n)-2,\frac 12 c+\frac 14 
n-1,\frac 12 d+\frac 14 n-1\right) }{\nabla ^{2}\left( \frac 12 b,
\frac 12 d+\frac 14 n-1,\frac 12 f+\frac 14 n-1\right) \Gamma \left(
e+\frac 12 n\right) }  \label{rcst}
\eea
(cf.\ (5.4) of \cite{Al87}), since some parameter from sets 
(\ref{sinta})--(\ref{sintc}) turns into 0 (possibly after using some 
symmetry property of 6$j $-symbols), together with fixed corresponding 
summation parameter, when other sum turns into summable balanced 
$_3F_2(1)$ series (see \cite{Sl66,GR90} and Appendix A of \cite{Al02a}). 
When vanishing linear combination of parameters of the stretched 
6$j$-symbol does not belong to sets (\ref{sinta})--(\ref{sintc}) the 
summation of special cases of (\ref{rinva})--(\ref{rinvc}) is more 
difficult.

In a near to the stretched case with $e=a+b-2$, the sum over $z_1$ in 
(\ref{rinva}) includes two terms and we have the $_4F_3(1)$ type sums 
over $z_2$ corresponding to the $6j$ coefficients of SU(2). Using for 
them the most symmetric (Racah) expression (see [29--31]) 
%\cite{JB77,VMK88,BL81}) 
we derive the following expression for special coefficients (\ref{cgr}): 
\bea
\fl c_{a,b,a+b-2;d,c,f}^{(\alpha ,n)}=\frac{(a,b,c,d,e,f)_{[n]}\Gamma 
\left( a+\frac 12 n-2\right) \Gamma \left( b+\frac 12 n-2\right) }{(n-3)!
\Gamma ^{3}\left( \frac 12 n\right) \nabla ^{2}\left( \frac 12 a,\frac 12
c+\frac 14 n-1,\frac 12 f+\frac 14 n-1\right) }  \nonumber \\
\times \frac{\nabla ^{2}\left( \frac 12 (e+n)-2,\frac 12 c+\frac 14 n-1,
\frac 12 d+\frac 14 n-1\right) }{64\Gamma \left( e+\frac 12 n+1\right) 
\nabla ^{2}\left( \frac 12 b,\frac 12 d+\frac 14 n-1,\frac 12 f+\frac 14 
n-1\right) }  \nonumber \\
\times \left\{ 2a(c+d-e)(e-c+d+n-2)\left[ (c+d-e+n-4)(b+d-f)\right.
\right.  \nonumber \\
\times \left. (a+c-f+n-4)-(c+d+e+2n-4)(a-c+f)(b-d+f)\right]  \nonumber \\
+(2e+n)(a-c+f)(c+f-a+n-2)\left[ (c+d+e+2n-4)\right.  \nonumber \\
\times (b-d+f)(a-c+f+n-4)-(b+d-f)(c+d-e)  \nonumber \\
\times \left. \left. (a+c-f+n-4)\right] \right\} .  \label{rbst}
\eea
Expressions (\ref{rcst}) and (\ref{rbst}) cover all but the last entrees 
of table 1 of \cite{HJu99}.

Otherwise, all three summation intervals in expression (\ref{rtr6j}) are
restricted by $r_{11}=\tfrac 12 (a+b-e)$. The sum over $g$ in 
(\ref{rtr6j}) turn into the very well-poised hypergeometric 
$_7F_6(1)$ series when we express the 6$j$-coefficients of SU(2) in 
the right-hand side of (\ref{rtr6j}) 
\bea
\fl \left\{ \begin{array}{ccc}
\frac 12 (b+n)-2 & \frac 12 d+\frac 14 n-1 & \frac 12 f+\frac 14 n-1 \\ 
\frac 12 c+\frac 14 n-1 & \frac 12 a & \frac 12 (g+n)-2
\end{array} \right\}  \nonumber \\
\times \left\{ \begin{array}{ccc}
\frac 12(g+n)-2 & \frac 12 b & \frac 12(a+n)-2 \\ 
\frac 12 f+\frac 14 n-1 & \frac 12 c+\frac 14 n-1 & \frac 12 d+\frac 14 
n-1
\end{array} \right\}  \label{ch6jt}
\eea
by means of (2.1a) (with inverted summation parameter) and (2.2a) of 
\cite{Al99,Al00}, respectively. Using Watson's transformation formula 
(2.5.1) of \cite{GR90} or (6.10) of \cite{LB94}, we rearrange the sum 
over $g$ into balanced $_4F_3(1)$ hypergeometric series (see also the 
related transition between expression (C3) for the $6j$ coefficients 
\cite{Al92} in terms of (\ref{nabla}) and expression (2.1a) of 
\cite{Al99,Al00}), with inverted sum. Then instead of (\ref{rtr6j}) we 
obtain the following triple sum expression for coefficients (\ref{cgr}): 
\bea
\fl c_{a,b,e;d,c,f}^{(\alpha ,n)}=(a,b,c,d,e,f)_{[n]}\frac{\left[ \frac 12
(a+b+e)+n-3\right] !\Gamma \left( \frac 12 (a+b-e+n)-1\right) }{%
(n-3)!\Gamma \left( \frac 12 (a+c+f+n)\right) \left[ \frac 12(a+c-f)
\right] !}  \nonumber \\
\times \frac{\Gamma \left( \frac 12(b+e-a+n)-1\right) \Gamma \left( 
\frac 12(a-b+e+n)-1\right) }{\Gamma ^{3}\left( \frac 12 n\right) \left[ 
\frac 12(b+e-a)\right] !\left[ \frac 12(a-b+e)\right] !}  \nonumber \\
\times \frac{\Gamma \left( \frac 12(d-b+f+n)-1\right) }{\left[ \frac 12
(b-d+f)\right] !}\sum_{z_1,z_2,z_3}\frac{(-1)^{(a+b-e)/2+z_1+z_2+z_3}
(a-z_1)!}{z_1!z_2!z_3!\left[ \frac 12 (c+d-a-b)+z_1\right] !}  \nonumber \\
\times \frac{\Gamma \left( \frac 12(c+f-a+n)-1+z_1\right) (b-z_2)!}{%
\left[ \frac 12(a-c+f)-z_1\right] !\left[ \frac 12(b+d-f)-z_2\right] !
(a+b+n-3-z_1-z_2)!}  \nonumber \\
\times \frac{\left[ \frac 12 (a+b-e)-z_3\right] !\Gamma \left( a+b+%
\frac 12(d-c-e+n)-1-z_1-z_2-z_3\right) }{\left[ \frac 12(a+b-e)-z_1
-z_3\right] !\left[ \frac 12 (a+b-e)-z_2-z_3\right] !\Gamma \left( e+
\frac 12 n+z_3\right) }  \nonumber \\
\times \frac{\left[ \frac 12 (a+b+c+d)+n-3-z_2\right] !\Gamma \left( 
\frac 12 (c-d+e+n)-1+z_3\right) }{\Gamma \left( \frac 12 (b+d+f+n)
-z_2\right) \Gamma \left( \frac 12(a+b-e+n)-1-z_3\right) } ,  \label{rin3}
\eea
where the separate sums are the balanced $_4F_3(1)$ series. The total
number of terms in (\ref{rin3}) does not exceed 
$\frac 16(r_{11}+1)(r_{11}+2)(2r_{11}+3)$ (but may be surpassed by 
$(r_{11}+1)(r_{13}+1)$ of (\ref{rinva}) or by $(r_{11}+1)(r_{31}+1)$ of 
(\ref{rinvb}) or (\ref{rinvc})). This expression does exhibit no usual or 
Regge type symmetry of $6j$-symbols of SO($n$).

It should be noted, that all three summation intervals for the triple sum 
expressions, derived directly from (\ref{df6j}) after used diverse 
expressions for the Racah coefficients (together with Dougall's summation 
formula \cite{Sl66,GR90} of $_5F_4(1)$ series) are never restricted by 
a single parameter.

\section{Expansions in terms of Pochhammer symbols and Kamp\'{e} de
F\'{e}riet series}

Using the parameters $r_{ik}=\beta _i-\alpha _{k}$ of modified 
Shelepin's \cite{Sh64} $R$-array (\ref{rasht}) (together with invariant 
parameters (\ref{sym6p})) and Pochhammer symbols, we may rewrite 
expressions for Regge symmetrical quantities in the following form: 
\numparts \bea
\fl \frac{c_{a,b,e;d,c,f}^{(\alpha ,n)}}{(a,b,c,d,e,f)_{[n]}}=\frac{%
(-1)^{\alpha _1-\alpha _3}(\alpha _3+n-3)!}{\Gamma ^{3}(n/2)(n-3)!r_{11}!
r_{12}!r_{13}!r_{14}!r_{21}!r_{33}!}  \nonumber \\
\times \Gamma \left[ {r_{22}+\tau ,r_{23}+\tau ,r_{24}+\tau ,r_{32}+\tau ,
r_{33}+\tau ,r_{34}+\tau }\atop {\alpha _2+\tau +1,\alpha _3+\tau +1,
\alpha _4+\tau +1}\right]  \nonumber \\
\times \sum_{x_1,x_2}\binom{r_{11}}{x_1}\binom{r_{13}}{x_2}%
(-1)^{x_1+x_2}(-r_{14},r_{22}+1,r_{23}+\tau )_{x_1}  \nonumber \\
\times (-r_{21},-\alpha _4-\tau ,r_{34}+\tau )_{r_{11}-x_1}  \nonumber \\
\times (r_{24}+\tau ,-r_{12}-\tau +1)_{x_2}(-\alpha _2-\tau ,r_{32}+
\tau )_{r_{13}-x_2}  \nonumber \\
\times (\beta _2-\beta _1+x_2+1)_{x_1}(-r_{21}-\tau -x_2+1)_{r_{11}-x_1} 
\label{rrva} \\
\lo= \frac{(-1)^{\beta _1-\beta _3}(\alpha _1+n-3)!}{\Gamma ^{3}(n/2)
(n-3)!r_{11}!r_{12}!r_{14}!r_{21}!r_{31}!r_{33}!}  \nonumber \\
\times \Gamma \left[ {r_{12}+\tau ,r_{22}+\tau ,r_{23}+\tau ,r_{24}+\tau ,
r_{33}+\tau ,r_{34}+\tau }\atop {\alpha _2+\tau +1,\alpha _3+\tau +1,
\alpha _4+\tau +1}\right]  \nonumber \\
\times \sum_{x_1,x_2}\binom{r_{11}}{x_1}\binom{r_{31}}{x_2}%
(-1)^{x_1+x_2}(-r_{14},r_{22}+1,r_{23}+\tau )_{x_1}  \nonumber \\
\times (-r_{21},r_{34}+\tau ,-\alpha _4-\tau )_{r_{11}-x_1}(-\alpha
_2-\tau ,-\alpha _3-n+3)_{x_2}  \nonumber \\
\times (r_{24}+\tau ,\alpha
_1+n-2)_{r_{31}-x_2}(r_{34}-x_2+1)_{r_{11}-x_1}  \nonumber \\
\times (-r_{34}-r_{11}-\tau +x_2+1)_{x_1}  \label{rrvb} \\
\lo= \frac{(-1)^{\beta _1-\beta _3}(\alpha _1+n-3)!}{\Gamma ^{3}(n/2)
(n-3)!r_{11}!r_{12}!r_{21}!r_{31}!r_{33}!r_{34}!}  \nonumber \\
\times \Gamma \left[ {r_{22}+\tau ,r_{23}+\tau ,r_{24}+\tau ,r_{32}+\tau ,
r_{33}+\tau ,r_{34}+\tau }\atop {\alpha _2+\tau +1,\alpha _3+\tau +1,
\alpha _4+\tau +1}\right]  \nonumber \\
\times \sum_{x_1,x_2}\binom{r_{11}}{x_1}\binom{r_{31}}{x_2}(-1)^{x_1+x_2}
(-r_{12},-\alpha _3-\tau ,-\alpha _4-\tau )_{x_1}  \nonumber \\
\times (r_{32}+\tau ,r_{22}+1,\alpha _1+n-2)_{r_{11}-x_1}  \nonumber \\
\times (r_{23}+\tau ,r_{24}+\tau )_{x_2}(-\alpha _2-\tau ,-r_{21}-\tau
+1)_{r_{31}-x_2}  \nonumber \\
\times (-r_{32}-r_{11}-\tau +x_2+1)_{x_1}(r_{32}-x_2+1)_{r_{11}-x_1},  
\label{rrvc}
\eea \endnumparts
where $\tau =n/2-1$. For products of several gamma functions in numerator
and denominator and Pochhammer symbols (appearing here only in the
numerator) we use the notations 
\begin{equation}
\Gamma \left[ {a_1,a_2,...,a_{A}}\atop {b_1,b_2,...,b_{B}}\right] =%
\frac{\Gamma (a_1)\Gamma (a_2)...\Gamma (a_{A})}{\Gamma (b_1)\Gamma
(b_2)...\Gamma (b_{B})},  \label{gamult}
\end{equation}
\begin{equation}
\fl (a_1,a_2,...,a_{A})_{k}=(a_1)_{k}(a_2)_{k}...(a_{A})_{k}=\frac{%
\Gamma (a_1+k)\Gamma (a_2+k)...\Gamma (a_{A}+k)}{\Gamma (a_1)\Gamma
(a_2)...\Gamma (a_{A})}.  \label{poch}
\end{equation}

Now we may express the terminating double hypergeometric series in 
(\ref{rrva})--(\ref{rrvc}) in terms of special Kamp\'{e} de F\'{e}riet 
\cite{K-F21} function $F_{1:3;3}^{1:4;4}[{{...}\atop {...}};1,1]$, which 
is defined as follows: 
\bea
\fl F_{1:3;3}^{1:4;4}\left[ {{a_1}\atop {c_1}}:{{(b)}\atop {(d)}};
{{(b^{\prime })}\atop {(d^{\prime })}};x,y\right]  \nonumber \\
\lo= \sum_{s,t}^{\infty }\frac{(a_1)_{s+t}}{s!t!(c_1)_{s+t}}\frac{%
\prod_{i=1}^{4}(b_i)_{s}(b_i^{\prime })_{t}}{\prod_{j=1}^{3}
(d_j)_{s}(d_j^{\prime })_{t}}x^{s}y^{t}.  \label{sKF}
\eea
and is terminating, because some separate numerator parameters are equal 
to negative integers: e.g., $b_i=-m$ and $b_{i^{\prime }}^{\prime }=-n$ 
in (\ref{sKF}), with $m$ and $n$ positive integers. In both cases some of 
the denominator parameters may be negative integers, but they should be 
smaller than the parameters responsible for the termination of series. 
Both separate series are balanced $_5F_4(1)$ series with parameters 
satisfying conditions 
\begin{equation}
\fl c_1-a_1=1+\sum_{i=1}^{4}b_i-\sum_{j=1}^{3}d_j=1+\sum_{i=1}^{4}
b_i^{\prime }-\sum_{j=1}^{3}d_j^{\prime }=\tau -1=n/2-2. 
\label{balc}
\end{equation}

Further we denote $r_{jk}+\tau $ by $\hat{r}_{jk}$, $\alpha _{k}
+\tau $ by $\hat{\alpha }_{k}$ and $\beta _j+\tau $ by $\hat{\beta}_j$ 
and write expression for (\ref{rrva}) in terms of special Kamp\'{e} de
F\'{e}riet series as follows: 
\numparts \bea
\fl \frac{c_{a,b,e;d,c,f}^{(\alpha ,n)}}{(a,b,c,d,e,f)_{[n]}}=
\frac{(\alpha _3+n-3)!}{\Gamma ^{3}(n/2)(n-3)!r_{11}!r_{12}!r_{13}!r_{14}!
r_{33}!(\beta _2-\beta _1)!}  \nonumber \\
\times \Gamma \left[ {\hat{r}_{21},\hat{r}_{22},\hat{r}_{23},\hat{r}_{24},
\hat{r}_{33},\hat{r}_{34}+r_{11},\hat{r}_{32}+r_{13}}\atop {\hat{\alpha}%
_3+1,\hat{\beta}_2-\beta _1,\hat{\alpha}_2-r_{13}+1,\hat{\alpha}_4-r_{11}
+1}\right]  \nonumber \\
\times F_{1:3;3}^{1:4;4}\left[ {{\beta _2-\beta _1+1}\atop {\hat{\beta}_2
-\beta _1}}:{{-r_{11},-r_{14},\hat{r}_{23},r_{22}+1}\atop {\beta
_2\!-\!\beta _1\!+\!1,\hat{\alpha}_4\!-\!r_{11}\!+\!1,-\hat{r}%
_{34}\!-\!r_{11}\!+\!1}};\right.  \nonumber \\
\qqquad \left. {{\hat{r}_{21},\hat{r}_{24},-r_{13},
-\hat{r}_{12}+1}\atop {\beta _2\!-\!\beta _1\!+\!1,-\hat{r}_{32}\!-\!%
r_{13}\!+\!1,\hat{\alpha}_2\!-\!r_{13}\!+\!1}};1,1\right]  \label{sKF1a} \\
\lo= \frac{(-1)^{\alpha _1-\alpha _3}(\alpha _3+n-3)!(r_{11}+r_{22})!
(r_{11}+r_{23})!}{\Gamma ^{3}(n/2)(n-3)!r_{11}!r_{12}!r_{13}!r_{21}!
r_{22}!r_{23}!r_{33}!(\alpha _1-\alpha _4)!}  \nonumber \\
\times \Gamma \left[ {\hat{r}_{12},\hat{r}_{22},\hat{r}_{32},\hat{r}%
_{33},\hat{r}_{34},r_{13}+\hat{r}_{24},r_{11}+\hat{r}_{23}}\atop {%
\hat{\alpha}_2+1,\hat{\alpha}_3+1,\hat{\alpha}_4+1,\hat{\alpha}_3
-\alpha _2}\right]  \nonumber \\
\times F_{1:3;3}^{1:4;4}\left[ {{-r_{11}\!-\!\hat{r}_{23}\!+\!1}\atop {%
-r_{11}-r_{23}}}:{{-r_{11},-r_{21},-\hat{\alpha}_4,\hat{r}_{34}}\atop {%
-r_{11}\!-\!\hat{r}_{23}\!+\!1,\alpha _1\!-\!\alpha
_4\!+\!1,-r_{11}\!-\!r_{22}}};\right.  \nonumber \\
\qqquad \left. {{-r_{23},-r_{13},\hat{r}_{32},
-\hat{\alpha}_2}\atop {-r_{11}\!-\!\hat{r}_{23}\!+\!1,-r_{13}\!-\!%
\hat{r}_{24}\!+\!1,\hat{\alpha}_3\!-\!\alpha _2}};1,1\right] , 
\label{sKF1b}
\eea \endnumparts
where parameters $r_{11}$ and $r_{13}$ of array (\ref{raSh}) are 
responsible for the termination of series in both cases (and may be 
replaced by $r_{21}$ and $r_{23}$ in the second (\ref{sKF1b}) case).

Similarly, we write expression for (\ref{rrvb}) in terms of special
Kamp\'{e} de F\'{e}riet series as follows:
\numparts \bea
\fl \frac{c_{a,b,e;d,c,f}^{(\alpha ,n)}}{(a,b,c,d,e,f)_{[n]}}=\frac{%
(-1)^{\beta _1-\beta _3}(r_{34}+r_{11})!(\beta _3+n-3)!}{\Gamma ^{3}
(n/2)(n-3)!r_{11}!r_{12}!r_{14}!r_{31}!r_{33}!r_{34}!(\beta _2
-\beta _1)!}  \nonumber \\
\times \Gamma \left[ {\hat{r}_{12},\hat{r}_{22},\hat{r}_{23},\hat{r}_{33},
\hat{r}_{24}+r_{31},\hat{r}_{34}+r_{11}}\atop {\hat{\alpha}_2+1,
\hat{\alpha}_3+1,\hat{\alpha}_4-r_{11}+1}\right]  \nonumber \\
\times F_{1:3;3}^{1:4;4}\left[ {{-\hat{r}_{34}\!-\!r_{11}\!+\!1}\atop {%
-r_{34}-r_{11}}}:{{-r_{11},-r_{14},\hat{r}_{23},r_{22}+1}\atop {-\hat{r}%
_{34}\!-\!r_{11}\!+\!1,\beta _2\!-\!\beta _1\!+\!1,\hat{\alpha}%
_4\!-\!r_{11}\!+\!1}};\right.  \nonumber \\
\qqquad \left. {{-r_{34},-r_{31},-\hat{\alpha}_2,-\alpha _3-n
+3}\atop {-\hat{r}_{14}\!-\!r_{31}\!+\!1,-\hat{r}_{24}\!-\!r_{31}\!+\!1,
-\beta _3\!-\!n\!+\!3}};1,1\right]  \label{sKF2a} \\
\lo= \frac{(\alpha _1+n-3)!(\alpha _3+n-3)!(r_{11}+r_{22})!}{\Gamma ^{3}
(n/2)(n-3)!r_{11}!r_{12}!r_{21}!r_{22}!r_{31}!r_{33}!(\alpha _3-r_{31}
+n-3)!}  \nonumber \\
\times \frac{1}{(\alpha _1-\alpha _4)!}\Gamma \left[ {\hat{r}_{12},
\hat{r}_{14},\hat{r}_{22},\hat{r}_{24},\hat{r}_{33},\hat{r}_{34},r_{11}
+\hat{r}_{23}}\atop {\hat{\alpha}_3+1,\hat{\alpha}_4+1,\hat{\alpha}_1-
\alpha _4,\hat{\alpha}_2-r_{31}+1}\right]  \nonumber \\
\times F_{1:3;3}^{1:4;4}\left[ {{\alpha _1-\alpha _4+1}\atop {\hat{%
\alpha}_1-\alpha _4}}:{{-r_{11},-r_{21},\hat{r}_{34},-\hat{\alpha}%
_4}\atop {\alpha _1\!-\!\alpha _4\!+\!1,-r_{11}\!-\!\hat{r}_{23}\!+\!1,
-r_{11}\!-\!r_{22}}};\right.  \nonumber \\
\qqquad \left. {{\hat{r}_{14},\hat{r}_{24},-r_{31},\alpha _1+
n-2}\atop {\alpha _1\!-\!\alpha _4\!+\!1,\hat{\alpha}_2\!-\!r_{31}\!+\!1,
\alpha _3\!-\!r_{31}\!+\!n\!-\!2}};1,1\right] ,  \label{sKF2b}
\eea \endnumparts
where parameters $r_{11}$ and $r_{31}$ are responsible for the 
termination of series in both cases (and may be replaced by $r_{14}$ and 
$r_{34}$ in the first (\ref{sKF2a}) case).

Finally, we write expression for (\ref{rrvc}) in terms of special 
Kamp\'{e} de F\'{e}riet series as follows: 
\numparts \bea
\fl \frac{c_{a,b,e;d,c,f}^{(\alpha ,n)}}{(a,b,c,d,e,f)_{[n]}}=
\frac{(-1)^{\beta _1-\beta _3}(\beta _1+n-3)!(r_{11}+r_{22})!(r_{11}
+r_{32})!}{\Gamma ^{3}(n/2)(n-3)!r_{11}!r_{12}!r_{21}!r_{22}!r_{31}!
r_{32}!r_{33}!r_{34}!}  \nonumber \\
\times \Gamma \left[ {\hat{r}_{21},\hat{r}_{22},\hat{r}_{23},\hat{r}_{24},
\hat{r}_{33},\hat{r}_{34},\hat{r}_{32}+r_{11}}\atop {\hat{\alpha}_3+1,
\hat{\alpha}_4+1,\hat{\alpha}_2-r_{31}+1,\hat{\beta}_2-\beta _3}\right] 
\nonumber \\
\times F_{1:3;3}^{1:4;4}\left[ {{-\hat{r}_{32}-r_{11}+1}\atop {-r_{32}-
r_{11}}}:{{-r_{11},-r_{12},-\hat{\alpha}_3,-\hat{\alpha}_4}\atop {-%
\hat{r}_{32}\!-\!r_{11}\!+\!1,-\beta _1\!-\!n\!+\!3,-r_{11}\!-\!r_{22}}}%
;\right.  \nonumber \\
\qqquad \left. {{-r_{32},-r_{31},\hat{r}_{24},\hat{r}_{23}%
}\atop {-\hat{r}_{32}\!-\!r_{11}\!+\!1,\hat{\alpha}_2\!-\!r_{31}\!+\!1,
\hat{\beta}_2\!-\!\beta _3}};1,1\right]  \label{sKF3a} \\
\lo= \frac{(\alpha _1+n-3)!}{\Gamma ^{3}(n/2)(n-3)!r_{11}!r_{21}!r_{31}!
r_{33}!r_{34}!(\alpha _1-\alpha _2)!}  \nonumber \\
\times \Gamma \left[ \frac{\hat{r}_{12},\hat{r}_{22},\hat{r}_{32},
\hat{r}_{33},\hat{r}_{34},\hat{r}_{23}+r_{31},\hat{r}_{24}+r_{31}}{%
\hat{\alpha}_2+1,\hat{\alpha}_3-r_{11}+1,\hat{\alpha}_4-r_{11}+1,
\hat{\alpha}_1-\alpha _2}\right]  \nonumber \\
\times F_{1:3;3}^{1:4;4}\left[ {{\alpha _1-\alpha _2+1}\atop {\hat{%
\alpha}_1-\alpha _2}}:{{-r_{11},\hat{r}_{32},\alpha _1+n-2,r_{22}+1%
}\atop {\alpha _1\!-\!\alpha _2\!+\!1,\hat{\alpha}_3\!-\!r_{11}\!+\!1,
\hat{\alpha}_4\!-\!r_{11}\!+\!1}};\right.  \nonumber \\
\qqquad \left. {{\hat{r}_{12},-r_{31},-\hat{\alpha}_2,
-\hat{r}_{21}+1}\atop {\alpha _1\!-\!\alpha _2\!+\!1,-\hat{r}_{24}\!-\!%
r_{31}\!+\!1,-\hat{r}_{23}\!-\!r_{31}\!+\!1}};1,1\right] ,  \label{sKF3b}
\eea \endnumparts
where parameters $r_{11}$ and $r_{31}$ are responsible for the 
termination of series in both cases (and may be replaced by $r_{12}$ and 
$r_{32}$ in the first (\ref{sKF3a}) case). However, the possible 
indefiniteness (appearing, e.g., with negative integer arguments of 
$(\beta _2-\beta _1)!$ or $\Gamma (\hat{\alpha}_1-\alpha _2)$ and never 
troubling in expressions (\ref{rrva})--(\ref{rrvc})) should be kept in 
attention in expressions (\ref{sKF1a})--(\ref{sKF3b}) in terms of special 
Kamp\'{e} de F\'{e}riet series.

There are the definite linear dependencies
\numparts \bea
a_1=d_1=d_1^{\prime }=d_j+d_j^{\prime }-1\;(j=2,3),  \nonumber \\
c_1=b_4+b_4^{\prime }-n+4=b_i+b_i^{\prime }\;(i=1,2,3)
\label{bala}
\eea
between parameters of each special Kamp\'{e} de F\'{e}riet series in 
(\ref{sKF1a}), (\ref{sKF2a}) and (\ref{sKF3b}), as well as the relations 
\bea
a_1=d_1=d_1^{\prime }=d_2+d_2^{\prime }-1=d_3+d_3^{\prime }-n+3,  
\nonumber \\
c_1=b_i+b_i^{\prime }\;(i=1,\cdot \cdot \cdot ,4)  \label{balb}
\eea \endnumparts
between parameters of each special Kamp\'{e} de F\'{e}riet series in 
(\ref{sKF1b}), (\ref{sKF2b}) and (\ref{sKF3a}). Of course, parameters 
$a_1=d_1=d_1^{\prime }$ (which are integers depending on the distance
between the rows or columns of Shelepin's array in (\ref{sKF1a}) or in 
(\ref{sKF2b}) and (\ref{sKF3b}), respectively) are never responsible for 
the termination of series. There is absent any correlation between the 
type of dependencies (\ref{bala})--(\ref{balb}) and the types of 
parameters $a_1$ and $c_1$, the last being non-positive integers in 
(\ref{sKF1b}), (\ref{sKF2a}) and (\ref{sKF3a}). Nevertheless, expressions 
(\ref{sKF1a}) and (\ref{sKF2a}) (as well as (\ref{sKF1b}) and 
(\ref{sKF2b})) are mutually related with respect to the substitution
(hook reflection)
\numparts \begin{equation}
d\rightarrow -d-n+2,  \label{hr1}
\end{equation}
leaving invariant dimension $d_{d}^{(n)}$ and character of irrep $d$ of 
SO($n$), when (\ref{sKF3b}) is invariant under this substitution, which 
corresponds to the transposition $\hat{r}_{32}\leftrightarrow \alpha _1
+n-2$ (together with $-r_{31}\leftrightarrow -\hat{\alpha}_2$) of 
parameters. Otherwise, expressions (\ref{sKF1a}) and (\ref{sKF3b}) (as 
well as (\ref{sKF1b}) and (\ref{sKF3a})) are mutually related with 
respect to the hook reflections
\begin{equation}
\fl c\rightarrow -c-n+2,\qquad d\rightarrow -d-n+2,\qquad 
f\rightarrow -f-n+2,  \label{hr3}
\end{equation} \endnumparts
leaving invariant dimensions and characters of irreps $c,\,d$ and $f$.

The transposition $r_{11}\leftrightarrow r_{14}$ (together with 
$\hat{r}_{24}\leftrightarrow \hat{r}_{21}$) in (\ref{sKF1a}) gives 
Regge symmetry (\ref{sym6Ra}), as well as $r_{11}\leftrightarrow r_{14}$ 
(together with $r_{31}\leftrightarrow r_{34}$) in (\ref{sKF2a}). The 
transposition $r_{11}\leftrightarrow r_{21}$ (together with 
$r_{13}\leftrightarrow r_{23}$) in (\ref{sKF1b}) gives Regge symmetry 
\begin{equation}
\left\{ \begin{array}{ccc}
a & b & e \\ 
d & c & f
\end{array} \right\} _{SO(n)}=\left\{ \begin{array}{ccc}
a & s_1-c & s_1-f \\ 
d & s_1-b & s_1-e
\end{array} \right\} _{SO(n)},  \label{sym6Rd}
\end{equation}
which corresponds to the transposition of rows in array (\ref{rasht}), as 
well as the transposition $r_{11}\leftrightarrow r_{21}$ (together with 
$\hat{r}_{14}\leftrightarrow \hat{r}_{24}$) in (\ref{sKF2b}). Otherwise,
transpositions $r_{11}\leftrightarrow r_{12}$ (together with 
$r_{31}\leftrightarrow r_{32}$) and $\hat{\alpha}_3\leftrightarrow 
\hat{\alpha}_4$ (together with $\hat{r}_{23}\leftrightarrow \hat{r}_{24}$)
are the generators of symmetries (\ref{sym6Rb}) and (\ref{sym6Rc}) for 
(\ref{sKF3a}).

Further, the transposition $-r_{14}\leftrightarrow \hat{r}_{23}$ (together
with $-r_{13}\leftrightarrow \hat{r}_{24}$ in (\ref{sKF1a}) or with 
$r_{31}\leftrightarrow \hat{\alpha}_{2}$ in (\ref{sKF2a}), respectively), 
the transpositions $r_{11}\leftrightarrow \hat{\alpha}_4$ (together 
with $-r_{23}\leftrightarrow \hat{r}_{32}$ or with $\hat{r}_{14}
\leftrightarrow \alpha _{1}+n-2$) and $-r_{21}\leftrightarrow 
\hat{r}_{34}$ (together with $r_{13}\leftrightarrow \hat{\alpha}_{2}$ or 
with $\hat{r}_{24}\leftrightarrow -r_{31}$) in (\ref{sKF1b}) or 
(\ref{sKF2b}), respectively, and the interchange of $r_{11}$ and 
$\hat{\alpha}_4$ (together with $-r_{32}\leftrightarrow \hat{r}_{23}$) 
or the interchange of $r_{12}$ and $\hat{\alpha}_{3}$ (together with 
$-r_{31}\leftrightarrow \hat{r}_{24}$) in (\ref{sKF3a}) correspond to 
some hook reflections. The remaining from 3! or 4! transpositions, 
leaving invariant dependencies (\ref{bala}) or (\ref{balb}), correspond 
to compositions of some Regge symmetries and hook reflections, as well 
as the interchanges of the sets $b_1,b_2,b_3,b_4;d_2,d_3$ and 
$b_1^{\prime },b_2^{\prime },b_3^{\prime },b_4^{\prime };d_2^{\prime },
d_3^{\prime }$ in (\ref{sKF}).

\section{Concluding remarks}

In this paper, we considered the recoupling coefficients and 
6$j$-symbols of the orthogonal SO($n$) groups for all six representations 
corresponding to the spherical or hyperspherical harmonics of these 
groups. Corrected triple sum expression of Ali\v sauskas \cite{Al87} 
(which was previously derived from the fourfold sum expression of 
\cite{Al87}, related to later expression of \cite{HJu99}) as an 
expansion in terms of the $6j$ coefficients of SU(2), with possible 
multiple of 1/4 parameters for odd $n$, here has been rearranged 
into three different double hypergeometric series of the Kamp\'{e} de 
F\'{e}riet $F_{1:3;3}^{1:4;4}$ type with the moderate (2$\times $2) 
symmetry which (together with hidden usual symmetry) nevertheless is 
resolving for all 144 Regge type symmetries. The different double sum 
expressions are mutually related with respect to the substitutions, 
generalizing the ``mirror reflection'' symmetry $j\rightarrow -j-1$ of 
the angular momentum theory \cite{JB77}. Note that more general double 
$F_{1:3;3}^{1:4;4}$ series (with balanced by condition (\ref{balc}) 
parameters and four (from 8) separate relations of the type (\ref{bala}) 
or (\ref{balb})) appeared as the doubly stretched $12j$ coefficients of 
the second kind \cite{Al02a} for SU(2) as presented by expressions 
(2.5a)--(2.5c) (with $z_1=z_2=0$) for $j_1=k_1-l_1=l_2-k_3$, or by 
expression (2.9) for $k_1=j_1+l_1=j_2+l_3$, although our 6$j$-symbols of 
SO($n$) cannot be reduced to special $12j$ coefficients of SU(2).

However, two non-positive integer parameters are necessary for 
termination of these double series, in contrast with special double 
hypergeometric series $F_{1:1,1}^{1:2,2}$ of the Kamp\'{e} de F\'{e}riet 
type \cite{Al00,LV-J01,V-JPR94} which correspond to the stretched $9j$ 
coefficients of SU(2), and may terminate for fixed single integer 
non-positive parameter, restricting all summation parameters. Single 
integer non-positive parameter is sufficient for restricting of all 
summation parameters only in the triple sum expression, derived in 
section 4. (This parameters appears in the double sums over $z_1,z_3$ and 
over $z_2,z_3$, which again may be treated as the double hypergeometric 
series $F_{1:2,2}^{1:3,3}$ of the Kamp\'{e} de F\'{e}riet type). 

Related recoupling coefficients of the symplectic groups Sp($2n$) with 
all six irreps antisymmetric are also given in Appendix as the double 
series which correspond to the orthogonal groups SO($-2n$) of negative 
dimension, with more rich restriction structure. These Sp($2n$) 
invariants never form the complete recoupling matrices, as well as the 
SO($n$) invariants considered in this paper as the main objects. Note 
the essential difference between our 6$j$-symbols and the 6$j$-symbols 
of the orthogonal SO($n$) groups for all six irreps antisymmetric and 
6$j$-symbols of the symplectic groups Sp($2n$) with all six irreps 
symmetric as expressed \cite{JLL90,JL87,CK82} in terms of the 
hypergeometric (single) $_4F_3(1)$ series with all 144 Regge type 
symmetries visible immediately. Alternating single sums appear in the 
Sp($2n$) case, but these $_4F_3(1)$ series are balanced 
(Saalsch\"{u}tzian) only in the Sp(2) or SU(2) case, when these 
6$j$-symbols also are forming the complete recoupling matrices. 

\appendix
\section{Recoupling coefficients for antisymmetric representations of 
Sp($2n$)}

As it was demonstrated by Judd \etal \cite{JLV00} the recoupling 
function\footnote{Which here is defined in a formal way with arbitrary 
labels of the basis states, without needed internal and external 
multiplicity labels.} 
\bea
\fl U\left( \begin{array}{ccc}
A & B & E \\ 
D & C & F
\end{array} \right) _{Sp(2n)}(AM_A,FM_F|CM_C)=\sum_{M_B,M_D,M_E}(DM_D,
BM_B|FM_F)  \nonumber \\
\times (AM_A,BM_B|EM_E)(EM_E,DM_D|CM_C)  \label{rcsp}
\eea
of the symplectic group Sp($2n$) for all six antisymmetric representations
$\langle 1^{\nu }\rangle $ ($\nu \leq n$), with dimension (cf. \cite{K72})
\begin{equation}
d_{\langle 1^{\nu }\rangle }^{\langle 2n\rangle }=
\frac{2(2n+1)!(n-\nu +1)}{\nu !(2n-\nu +2)!}  \label{dimsp}
\end{equation}
may be expressed as an analytical continuation of the recoupling 
coefficients for symmetric representations of the orthogonal group of the 
negative rank SO($-2n$), in accordance with \cite{JLL90,JLV00,CK82} 
(cf.\ also \cite{M80,D89}) as follows: 
\begin{equation}
\fl U\left( \begin{array}{ccc}
\langle 1^{a}\rangle & \langle 1^{b}\rangle & 
\langle 1^{e}\rangle \\ 
\langle 1^{d}\rangle & \langle 1^{c}\rangle & 
\langle 1^{f}\rangle 
\end{array}
\right) _{Sp(2n)}=(-1)^{\chi _{12}}\left( d_e^{(-2n)}d_f^{(-2n)}
\right) ^{1/2}\left\{ \begin{array}{ccc}
a & b & e \\ 
d & c & f
\end{array} \right\} _{SO(-2n)},  \label{rsp6j}
\end{equation}
(ignoring possible phase factor $(-1)^{\chi }$).

It is most convenient to perform the analytical continuation of 
expressions (\ref{rrva})--(\ref{rrvc}), together with (\ref{rd6jb}) and 
(\ref{nabls}), using the same integer parameters $r_{ik}=\beta _i-
\alpha _{k}$ ($i=1,2,3;\;j=1,\cdot \cdot \cdot ,4$) of array 
(\ref{rasht}), or the most symmetric parametrization (\ref{sym6p}). 
Hence replacing $n$ by $-2n$ and $\tau $ by $-n-1$ we write the following 
three expressions for the recoupling coefficients of the symplectic group:
\begin{eqnarray*}
\fl U\left( \begin{array}{ccc}
\langle 1^{a}\rangle  & \langle 1^{b}\rangle  & 
\langle 1^{e}\rangle  \\ 
\langle 1^{d}\rangle  & \langle 1^{c}\rangle  & 
\langle 1^{f}\rangle 
\end{array}
\right) _{Sp(2n)}=\frac{(-1)^{\chi _{12}}\left( d_{\langle 1^{e}\rangle %
}^{\langle 2n\rangle }d_{\langle 1^{f}\rangle }^{\langle 2n\rangle %
}\right) ^{1/2}N n!^{3}(2n+2)!}{(2n+2-\alpha _3)!r_{11}!r_{12}!r_{13}!
r_{14}!r_{21}!r_{33}!(n-r_{22}+1)!} \\
\times \frac{(n-\alpha _2)!(n-\alpha _3)!(n-\alpha _4)!}{(n-r_{23}+1)!
(n-r_{24}+1)!(n-r_{32}+1)!(n-r_{33}+1)!(n-r_{34}+1)!} \\
\times \sum_{x_1,x_2}\binom{r_{11}}{x_1}\binom{r_{13}}{x_2}%
(-1)^{x_1+x_2}(-r_{14},r_{22}+1,r_{23}-n-1)_{x_1} \\
\times (-r_{21},-\alpha _4+n+1,r_{34}-n-1)_{r_{11}-x_1} \\
\times (r_{24}-n-1,-r_{12}+n+2)_{x_2}(-\alpha _2+n+1,r_{32}-n-1)_{r_{13}
-x_2} \\
\times (\beta _2-\beta _1+x_2+1)_{x_1}(-r_{21}+n+2-x_2)_{r_{11}-x_1}  
\qqquad \qquad (A.4a) \\ %\label{rcspa} 
\lo= \frac{(-1)^{\chi _{12}}\left( d_{\langle 1^{e}\rangle }^{\langle 2n
\rangle }d_{\langle 1^{f}\rangle }^{\langle 2n\rangle }\right) ^{1/2}N 
n!^{3}(2n+2)!}{(2n-\alpha _1+2)!r_{11}!r_{12}!r_{14}!r_{21}!r_{31}!r_{33}!
(n-r_{12}+1)!(n-r_{22}+1)!} \\
\times \frac{(n-\alpha _2)!(n-\alpha _3)!(n-\alpha _4)!}{(n-r_{23}+1)!
(n-r_{24}+1)!(n-r_{33}+1)!(n-r_{34}+1)!} \\
\times \sum_{x_1,x_2}\binom{r_{11}}{x_1}\binom{r_{31}}{x_2}(-1)^{x_1+x_2}
(-r_{14},r_{22}+1,r_{23}-n-1)_{x_1} \\
\times (-r_{21},r_{34}-n-1,-\alpha _4+n+1)_{r_{11}-x_1}(-\alpha _2+n+1,
-\alpha _3+2n+3)_{x_2} \\
\times (r_{24}-n-1,\alpha _1-2n-2)_{r_{31}-x_2}(r_{34}-x_2+1)_{r_{11}-x_1} 
\\
\times (-r_{34}-r_{11}+n+x_2+2)_{x_1}  \qqquad \qqquad \qquad \quad (A.4b) 
\\ %\label{rcspb} \\
\lo= \frac{(-1)^{\chi _{12}+\beta _1-\beta _3}\left( d_{\langle 1^{e}
\rangle }^{\langle 2n\rangle }d_{\langle 1^{f}\rangle }^{\langle 2n
\rangle }\right) ^{1/2}N n!^{3}(2n+2)!}{(2n-\alpha _1+2)!r_{11}!r_{12}!
r_{21}!r_{31}!r_{33}!r_{34}!(n-r_{22}+1)!} \\
\times \frac{(n-\alpha _2)!(n-\alpha _3)!(n-\alpha _4)!}{(n-r_{23}+1)!
(n-r_{24}+1)!(n-r_{33}+1)!(n-r_{34}+1)!} \\
\times \sum_{x_1,x_2}\binom{r_{11}}{x_1}\binom{r_{31}}{x_2}(-1)^{x_1+x_2}
(-r_{12},-\alpha _3+n+1,-\alpha _4+n+1)_{x_1} \\
\times (r_{32}-n-1,r_{22}+1,\alpha _1-2n-2)_{r_{11}-x_1} \\
\times (r_{23}-n-1,r_{24}-n-1)_{x_2}(-\alpha _2+n+1,-r_{21}+n+2)_{r_{31}
-x_2} \\
\times (-r_{32}-r_{11}+n+x_2+2)_{x_1}(r_{32}-x_2+1)_{r_{11}-x_1},
\qqquad \quad (A.4c) %\label{rcspc}
\end{eqnarray*}
where actually some ratios of gamma functions $\Gamma (-y)/\Gamma (-x)$
turned into the ratios of factorials $(-1)^{x-y}x!/y!$ only in the factor 
under the square root of (\ref{rd6jb}) (together with (\ref{nabls})) 
which absolute value appeared as 
\[
N=\frac{\left[ \prod_{i=1}^{3}\prod_{k=1}^{4}r_{ik}!(n+1-r_{ik})!\right]
^{1/2}}{(2n+2)!^{2}n!^{4}}\left[ \prod_{k=1}^{4}\frac{(2n+2-
\alpha _{k})!}{(n-\alpha _{k})!}\right] ^{1/2},
\]
in the dimension factors and in the factors before the summation sign of 
(\ref{rrva})--(\ref{rrvc}). Intervals of summation in (A.4$a$)--(A.4$c$)
%(\ref{rcspa})--(\ref{rcspc}) 
are restricted not only by conditions (\ref{sinta})--(\ref{sinte}) 
(e.g., $r_{11}\geq 0$ and $r_{13}\geq 0,$ or $r_{31}\geq 0$)
but also by conditions of the type $n+1-r_{ik}\geq 0$. Therefore it may 
be convenient to write (A.4$a$)--(A.4$c$) %(\ref{rcspa})--(\ref{rcspc}) 
as factorial series, which summation intervals are determined by 
non-negative arguments of the denominator factorials. Some factorials 
before the sum signs cancel. All the separate sums in such new version 
of (A.4$a$)--(A.4$c$) %(\ref{rcspa})--(\ref{rcspc}) 
are alternating, with exception of the sum over $x_1$ in (A.4$c$).
%(\ref{rcspc}). 
Of course, the recoupling coefficients of the symplectic group Sp($2n$) 
with all six antisymmetric irreps vanish unless $n-\alpha _{k}\geq 0$. 
The number of different recoupling coefficients of Sp($2n$) for fixed $n$ 
is finite, in contrast with the number of 6$j$-symbols of SO($n$), 
considered in this paper. 

The symmetry properties of the recoupling functions of the symplectic group
Sp($2n$) may be more complicated, e.g.
\begin{eqnarray*}
\fl \left( d_{\langle 1^{b}\rangle }^{\langle 2n\rangle }
d_{\langle 1^{c}\rangle }^{\langle 2n\rangle }\right) ^{-1/2}U\left( 
\begin{array}{ccc}
\langle 1^{a}\rangle  & \langle 1^{e}\rangle  & \langle 1^{b}\rangle  \\ 
\langle 1^{d}\rangle  & \langle 1^{f}\rangle  & \langle 1^{c}\rangle 
\end{array} \right) _{Sp(2n)}= (-1)^{\chi _{13}-\chi _{12}+(b+c-e-f)/2}
\\
\times \left( d_{\langle 1^{e}\rangle }^{\langle 2n\rangle }
d_{\langle 1^{f}\rangle }^{\langle 2n\rangle }\right) ^{-1/2}
U\left( \begin{array}{ccc}
\langle 1^{a}\rangle  & \langle 1^{b}\rangle  & \langle 1^{e}\rangle  \\ 
\langle 1^{d}\rangle  & \langle 1^{c}\rangle  & \langle 1^{f}\rangle 
\end{array} \right) _{Sp(2n)}.  \qqquad \qquad (A.5) %\label{sym6sp}
\end{eqnarray*}
The 6$j$-symbols of Sp($2n$) in the case of all antisymmetric irreps are
real and invariant under all permutations of the type (\ref{sym6j}), if 
we choose $\chi _{13}=\beta _2$, $\chi _{12}=\beta _1$ as defined by 
(\ref{sym6p}).

\end{document}